\documentclass[conference]{IEEEtran}

\DeclareMathAlphabet{\mathitbf}{OML}{cmm}{b}{it}

\usepackage{tikz}
\usetikzlibrary{matrix,arrows}

\usepackage{amsmath,amssymb,mathrsfs,amsthm}
\usepackage[colorlinks=true,linkcolor=black,anchorcolor=black,citecolor=black,filecolor=black,menucolor=black,urlcolor=black]{hyperref}
\usepackage{graphicx}
\usepackage{psfrag}
\usepackage{subfigure}
\usepackage{cite}
\usepackage{multirow}
\usepackage{stfloats}
\usepackage[margin=1in]{geometry}
\usepackage{gensymb}

\usepackage{epsfig}
\usepackage{amsfonts}
\usepackage{graphics}
\usepackage{color}
\usepackage{mathtools}
 
\DeclarePairedDelimiter{\floor}{\lfloor}{\rfloor} 
\epsfverbosetrue

\usepackage{xspace}
\usepackage{bbm}
\catcode`@=11
\chardef\mathlig@atcode\count255

\def\actively#1#2{\begingroup\uccode`\~=`#2\relax\uppercase{\endgroup#1~}}
\def\mathlig@gobble{\afterassignment\mathlig@next@cmd\let\mathlig@next= }
\def\mathlig@delim{\mathlig@delim}
\def\mathlig@defcs#1{\expandafter\def\csname#1\endcsname}
\def\mathlig@let@cs#1#2{\expandafter\let\expandafter#1\csname#2\endcsname}
\def\mathlig@appendcs#1#2{\expandafter\edef\csname#1\endcsname{\csname#1\endcsname#2}}

\def\mathlig#1#2{\mathlig@checklig#1\mathlig@end\mathlig@defcs{mathlig@back@#1}{#2}\ignorespaces}

\def\mathlig@checklig#1#2\mathlig@end{%
 \expandafter\ifx\csname mathlig@forw@#1\endcsname\relax
 \expandafter\mathchardef\csname mathlig@back@#1\endcsname=\mathcode`#1%
 \mathcode`#1"8000\actively\def#1{\csname mathlig@look@#1\endcsname}%
 \mathlig@dolig#1\mathlig@delim
\fi
\mathlig@checksuffix#1#2\mathlig@end
}

\def\mathlig@checksuffix#1#2\mathlig@end{%
\ifx\mathlig@delim#2\mathlig@delim\relax\else\mathlig@checksuffix@{#1}#2\mathlig@end\fi
}
\def\mathlig@checksuffix@#1#2#3\mathlig@end{%
\expandafter\ifx\csname mathlig@forw@#1#2\endcsname\relax\mathlig@dosuffix{#1}{#2}\fi
\mathlig@checksuffix{#1#2}#3\mathlig@end
}

\def\mathlig@dosuffix#1#2{%
\mathlig@appendcs{mathlig@toks@#1}{#2}%
\mathlig@dolig{#1}{#2}\mathlig@delim
}

\def\mathlig@dolig#1#2\mathlig@delim{%
 \mathlig@defcs{mathlig@look@#1#2}{%
 \mathlig@let@cs\mathlig@next{mathlig@forw@#1#2}\futurelet\mathlig@next@tok\mathlig@next}%
 \mathlig@defcs{mathlig@forw@#1#2}{%
  \mathlig@let@cs\mathlig@next{mathlig@back@#1#2}%
  \mathlig@let@cs\checker{mathlig@chck@#1#2}%
  \mathlig@let@cs\mathligtoks{mathlig@toks@#1#2}%
  \expandafter\ifx\expandafter\mathlig@delim\mathligtoks\mathlig@delim\relax\else
  \expandafter\checker\mathligtoks\mathlig@delim\fi
  \mathlig@next
 }%
 \mathlig@defcs{mathlig@toks@#1#2}{}%
 \mathlig@defcs{mathlig@chck@#1#2}##1##2\mathlig@delim{%
  \ifx\mathlig@next@tok##1%
   \mathlig@let@cs\mathlig@next@cmd{mathlig@look@#1#2##1}\let\mathlig@next\mathlig@gobble
  \fi 
  \ifx\mathlig@delim##2\mathlig@delim\relax\else
   \csname mathlig@chck@#1#2\endcsname##2\mathlig@delim
  \fi
 }%
 \ifx\mathlig@delim#2\mathlig@delim\else
  \mathlig@defcs{mathlig@back@#1#2}{\csname mathlig@back@#1\endcsname #2}%
 \fi
}%

\catcode`@\mathlig@atcode

\newcommand{\muspace}{\mspace{1mu}}

\DeclareRobustCommand{\scond}{\mathchoice{\muspace\vert\muspace}{\vert}{\vert}{\vert}}
\mathlig{|}{\scond}

\DeclareRobustCommand{\discint}{\mathchoice{\mspace{-1.5mu}:\mspace{-1.5mu}}{\mspace{-1.5mu}:\mspace{-1.5mu}}{:}{:}}
\mathlig{::}{\discint}
\newcommand{\suchthat}{\mathchoice{\colon}{\colon}{:\mspace{1mu}}{:}}

\newcommand{\Cr}{\mathscr{C}}
\newcommand{\Crn}{\mathscr{C}^{\degree}}

\newcommand{\Rr}{\mathscr{R}}

\newcommand{\Rv}{{\bf R}}

\newcommand{\tv}{{\bf t}}

\newcommand{\sv}{{\bf s}}

\newcommand{\lambdav}{{\boldsymbol \lambda}}
\newcommand{\muv}{{\boldsymbol \mu}}

\def\G{\Gamma}

\def\textiid{i.i.d.\@\xspace}
\newcommand\iid{\ifmmode\text{ i.i.d. } \else \textiid \fi}

\def\mathllap{\mathpalette\mathllapinternal}
\def\mathllapinternal#1#2{%
  \llap{$\mathsurround=0pt#1{#2}$}}

\def\clap#1{\hbox to 0pt{\hss#1\hss}}
\def\mathclap{\mathpalette\mathclapinternal}
\def\mathclapinternal#1#2{%
  \clap{$\mathsurround=0pt#1{#2}$}}

\let\oldstackrel\stackrel
\renewcommand{\stackrel}[2]{\oldstackrel{\mathclap{#1}}{#2}}

\renewcommand{\hbar}{h\mathllap{\overline{\vphantom{h}\hphantom{\rule{4.6pt}{0pt}}}\mspace{0.77mu}}}

\catcode`~=11 
\newcommand{\urltilde}{\kern -.06em\lower -.06em\hbox{~}\kern .02em}
\catcode`~=13 

\newcommand{\csym}{C_\mathrm{sym}}
\newcommand{\rsym}{R_\mathrm{sym}}
\newcommand{\csum}{C_\mathrm{sum}}
\newcommand{\rsum}{R_\mathrm{sum}}
\newcommand{\ov}{{\bf{1}}}

\def\0{\bf{0}}

\newtheorem{theorem}{Theorem}
\newtheorem{lemma}{Lemma}
\newtheorem{proposition}{Proposition}

\theoremstyle{definition}
\newtheorem{definition}{Definition}

\newtheorem{corollary}{Corollary}

\newtheorem{remark}{Remark}

\begin{document}

\title{Three Stories on a Two-sided Coin:\\
Index Coding, Locally Recoverable Distributed Storage, and Guessing Games on Graphs}
\author{
\IEEEauthorblockN{Fatemeh Arbabjolfaei and Young-Han Kim}
\IEEEauthorblockA{Department of Electrical and Computer Engineering\\
University of California, San Diego\\
Email: \{farbabjo, yhk\}@ucsd.edu
}
}
\date{}
\maketitle

\begin{abstract}
Three science and engineering
problems of recent interests---index coding, locally recoverable distributed storage, and guessing games
on graphs---are discussed and the connection between their optimal solutions is elucidated.
By generalizing recent results by Shanmugam and Dimakis and by Mazumdar on the complementarity between the optimal
broadcast rate of an index coding problem on a directed graph 
and the normalized rate of a locally recoverable distributed storage problem 
on the same graph, it is shown that
the capacity region and the optimal rate region of these two problems are complementary.
The main ingredients in establishing this result are the notion of confusion graph introduced by Alon et al. (2008),
the vertex transitivity of a confusion graph, the characterization of the index coding capacity region
via the fractional chromatic number of confusion graphs, and the characterization of the optimal rate region of the locally recoverable distributed storage
via the independence number of  confusion graphs.
As the third and final facet of the complementarity,
guessing games on graphs by Riis are discussed as special cases of the locally recoverable distributed storage problem,
and it is shown that the winning probability of the optimal strategy for a guessing game and 
the ratio between the winning probabilities of the optimal strategy and a random guess can be characterized, respectively,
by the capacity region for index coding and the optimal rate region for distributed storage.
\end{abstract}

\section{Introduction}

This paper discusses index coding, locally recoverable distributed storage, and guessing games on directed graphs,
with the goal of elucidating
the relationship between the optimal solutions of these three problems.

\emph{Index coding.} 
A server has multiple messages and wishes to send them to their corresponding receivers.
Each receiver has some side information about a subset of messages (not including the desired message).
The goal is to exploit the  side information of the receivers to minimize the number of required transmissions.
This problem arises in many contexts such as intra-cell communication, content broadcasting and coded caching \cite{Maddah-Ali--Niesen2014},
as well as satellite communication as originally introduced by Birk and Kol \cite{Birk--Kol2006}.

It is easy to see that index coding is a special case of network coding which is a key open problem in network information theory.
Effros, El Rouayheb, and Langberg \cite{Effros--El-Rouayheb--Langberg2013} showed that 
for any network coding problem, there exists an equivalent index coding problem.
This equivalence makes index coding even more intriguing.
So far the index coding problem has been attacked using tools from various disciplines such as graph theory \cite{Birk--Kol2006, Alon--Lubetzky--Stav--Weinstein--Hassidim2008, Blasiak--Kleinberg--Lubetzky2013, Shanmugam--Dimakis--Langberg2013b, Tahmasbi--Shahrasbi--Gohari2014b, Arbabjolfaei--Kim2015, Yi--Sun--Jafar--Gesbert2015, Ong--Lim--Ho2013}, algebra \cite{Bar-Yossef--Birk--Jayram--Kol2011, Huang--El-Rouayheb2015}, interference alignment \cite{Maleki--Cadambe--Jafar2014, Jafar2014, Sun--Jafar2015}, source coding \cite{Unal--Wagner2014}, and random coding \cite{Arbabjolfaei--Bandemer--Kim--Sasoglu--Wang2013}. 

\emph{Locally recoverable distributed storage.} A set of servers collectively store data such that 
if a server fails, its contents can be efficiently reconstructed from the contents of the other servers (among many others, see \cite{Shah--Rashmi--Kumar--Ramchandran2012, Papailiopoulos--Dimakis--Cadambe2013, Cadambe--Jafar--Maleki--Ramchandran--Suh2013}). 
The goal is to design a distributed storage code that maximizes the amount of data that can be stored while satisfying the  single-failure recovery constraint.
In the mentioned references, it is assumed that each server is able to connect to all the other servers of the system.
However, in a real system, due to some constraints, a server may only have access to a subset of the other servers.
In \cite{Mazumdar2014}, Mazumdar took this point into consideration by assuming that the topology of the system is given by a directed graph  and established a duality between the normalized rate of such a distributed storage system and the broadcast rate of the index coding problem.
In particular, Mazumdar showed that for any directed graph on $n$ nodes, the broadcast rate of the index coding problem and the normalized  rate of the locally recoverable distributed storage problem sum up to $n$.
In an independent concurrent work, Shanmugam and Dimakis \cite{Shanmugam--Dimakis2014} established a similar result for vector linear codes and showed that the dual code of a linear index code is a linear code for the locally recoverable distributed storage problem and vice versa.

\emph{Guessing games on graphs.}
Consider a cooperative game among multiple players, in which a value (hat color) is assigned to each player independently.
The value of each player is not revealed to her, but she knows the values assigned to a subset of the other players she can observe.
The players should simultaneously guess their own values and they win if all of them guess correctly.
After the values are assigned, the players are not allowed to communicate, but they can come up with a strategy beforehand to maximize their probability of winning. What is the best strategy and how much improvement over random guesses
can we make by employing the optimal strategy?
This interesting puzzle, which apparently seems to be in the realm of recreational mathematics,
was introduced originally by Riss \cite{Riis2007} for studying multiple unicast network coding problems.
Its relationship to index coding was observed firstly by Yi, Sun, Jafar, and Gesbert \cite{Yi--Sun--Jafar--Gesbert2015}.

\begin{figure}
\begin{center}
\footnotesize
\psfrag{ic}[c]{Index coding}
\psfrag{ds}[c]{Distributed storage}
\psfrag{gg}[c]{Optimal guessing number}
\psfrag{gg2}[c]{guessing number}
\psfrag{ggc}[c]{Optimal complementary}
\includegraphics[scale=0.4]{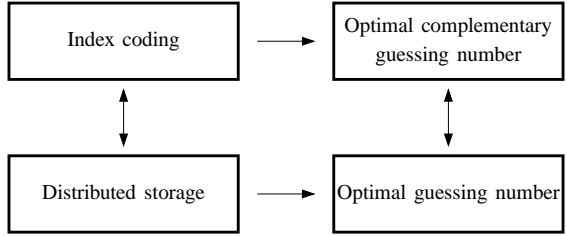}
\end{center}
\caption{The relationship between index coding, distributed storage, optimal guessing number, and optimal complementary guessing number.}
\label{fig:icvsds}
\end{figure}

In this paper, we define the $\lambdav$-directed capacity $C(\lambdav)$ of the index coding problem and the $\lambdav$-directed optimal rate $R(\lambdav)$ of the distributed storage problem and rewrite the capacity region of the index coding problem and the optimal rate region of the distributed storage problem in terms of the set of $\lambdav$-directed capacities, and the set of $\lambdav$-directed optimal rates respectively.
Denoting the all-ones vector by $\ov$, $C(\ov)$ is equal to the reciprocal of the broadcast rate of the index coding problem and  $R(\ov)$ equals the reciprocal of the normalized rate of the distributed storage.
The complementarity  between the capacity region and the optimal rate region of the two problems is demonstrated by generalizing the complementarity result by Mazumdar \cite{Mazumdar2014}, and by Shanmugam and Dimakis \cite{Shanmugam--Dimakis2014}, to a complementarity relationship between $C(\lambdav)$ and $R(\lambdav)$ for any nonnegative real $n$-tuple $\lambdav$.
The same technique is used to make a complementary connection between the sum-capacity of the index coding problem and the optimal sum-rate of the distributed storage.
Next, we will discuss that the optimal guessing number of guessing games on graphs is inversely related to the optimal sum-rate of the distributed storage problem which together with the complementary connection between the sum-capacity and the optimal sum-rate makes the inverse relationship between the optimal complementary guessing number and the sum-capacity of the index coding problem clear (see Fig. \ref{fig:icvsds}).
As a result of the clear connection between the optimal solutions of these three problems, the known results for one of these problems can be easily extended to the other problems. We will conclude the paper by one such example.

The rest of the paper is organized as follows.
We first review some mathematical preliminaries in Section \ref{sec:prelim}.
In Section \ref{sec:def}, formal definitions of the problems are presented.
In Section \ref{sec:capacity}, we overview the characterization of the capacity region of the index coding problem, the optimal rate region of the distributed storage problem, and the guessing number of the guessing games via the notion of confusion graph.
Section \ref{sec:duality} is the main part of the paper in which the complementarity between the index coding problem  and the distributed storage problem is presented and the guessing game on a graph is shown to be a special case of the distributed storage problem on the same graph.
Finally, in Section \ref{sec:conclusion}, we conclude by mentioning some results for the guessing games that due to the established connections, also hold for the index coding problem.

\section{Mathematical Preliminaries}
\label{sec:prelim}

Throughout the paper, $\ov$ denotes the $n \times 1$ vector of all ones.
The notations $V(\G)$ and $E(\G)$ mean the vertex set and the edge set of a graph $\G$ respectively.
For $v_1, v_2 \in V(\G)$, $v_1 \sim v_2$ denotes that there exists an edge between $v_1$ and $v_2$.
The set of positive integers that are less than or equal to $n$, $\{1,2,\ldots, n\}$, is denoted by $[n]$.

\subsection{Graph Coloring}

A (vertex) coloring of an undirected graph $\G$ 
is a mapping that assigns a color to each vertex
such that no two adjacent vertices share the same color.
The \emph{chromatic number} $\chi(\G)$ is the minimum number of colors such that a coloring of the graph exists.
More generally, a $b$-fold coloring assigns a set of $b$ colors to each vertex 
such that no two adjacent vertices share the same color.
The $b$-fold chromatic number $\chi^{(b)}(\G)$ is the minimum number of colors such that a $b$-fold coloring
exists.
The \emph{fractional chromatic number} of the graph is defined as%
\[
\chi_f(\G) = \lim_{b \rightarrow \infty} \frac{\chi^{(b)}(\G)}{b} = \inf_b \frac{\chi^{(b)}(\G)}{b},
\]
where the limit exists since $\chi^{(b)}(\G)$ is subadditive.
Consequently,
$\chi_f(\G) \le \chi(\G)$.

\subsection{Vertex Transitive Graphs}

\begin{definition}
An automorphism of a graph $\G$ is a 
bijective function $\sigma: V(\G) \to V(\G)$, such that the pair of vertices $(\sigma(u),\sigma(v))$ form an edge iff (if and only if) the pair $(u,v)$ form an edge. 
\end{definition}

\begin{definition}
A graph $\G$ is said to be vertex transitive if for any two vertices $u$ and $v$ of $\G$, there exists some automorphism  $\sigma: V(\G) \to V(\G)$ such that $\sigma(u) = v$. 
\end{definition}

\begin{definition}
The independence number  of the undirected graph $\G$ denoted $\alpha(\G)$ is the cardinality of the largest independent set of the graph.
\end{definition}

\begin{lemma}[Scheinerman and Ullman {\cite[Prop.~3.1.1]{Scheinerman--Ullman2011}}]
\label{lem:vertextransitive}
For any vertex transitive graph $\G$ we have
\[
\chi_f(\G) = |V(\G)|/\alpha(\G).
\]
\end{lemma}

\subsection{Disjunctive Graph Product}

Given two (undirected) graphs $\G_1$ and $\G_2$, the disjunctive product $\G = \G_1*\G_2$ produces a graph $\G$ on
the Cartesian product of the original vertex sets, i.e., $V(\G) = V(\G_1) \times V(\G_2)$ and the edge set is constructed from
the original edge sets according to the following rule. 
$(u_1,u_2) \sim (v_1,v_2)$ iff
$u_1 \sim v_1$ or $u_2 \sim v_2$.
The fractional chromatic number of the disjunctive product is multiplicative.

\begin{lemma}[Scheinerman and Ullman {\cite[Cor.~3.4.2]{Scheinerman--Ullman2011}}]
\label{lem:orcoloring}
\[ 
\chi_f(\G_1*\G_2) = \chi_f(\G_1)\chi_f(\G_2). 
\]
\end{lemma}

\subsection{Confusion Graphs}

The notion of confusion graph is originally introduced by Alon, Hassidim, Lubetzky, Stav, and Weinstein \cite{Alon--Hassidim--Lubetzky--Stav--Weinstein2008}.
Consider a directed graph $G=(V,E)$ with $V = [n]$. Let $A_j = \{i \in V \suchthat (i,j) \in E\}$, $j \in [n]$ and let
$\tv = (t_1, \ldots, t_n)$ be a length-$n$ integer tuple.
Two binary $n$-tuples $x^n, z^n \in \prod_{i=1}^n \{0,1\}^{t_i}$ are said to be \emph{confusable
at node $j \in [n]$} if $x_j \ne z_j$ and $x_i = z_i$ for all $i \in A_j$.
We say $x^n$ and $z^n$ are \emph{confusable} if they are confusable at some node ~$j$.

\begin{figure}
\vspace{0.75em}
\begin{center}
\subfigure[]{
\small
\psfrag{1}[cb]{1}
\psfrag{2}[rc]{2}
\psfrag{3}[lc]{3}
\psfrag{4}{4}
\includegraphics[scale=0.45]{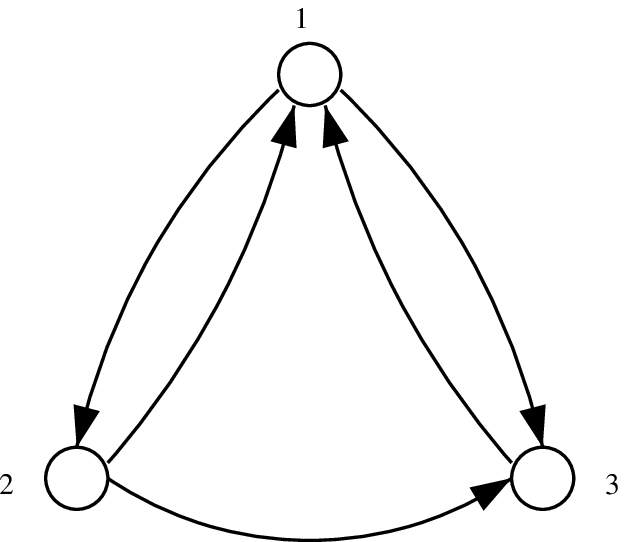}
}
\subfigure[]{
\vspace{0.5em}
\small
\psfrag{000}[cl]{000}
\psfrag{001}[cl]{001}
\psfrag{010}[l]{010}
\psfrag{011}[l]{011}
\psfrag{100}[clb]{100}
\psfrag{101}[clb]{101}
\psfrag{110}[]{110}
\psfrag{111}[]{111}
\includegraphics[scale=0.45]{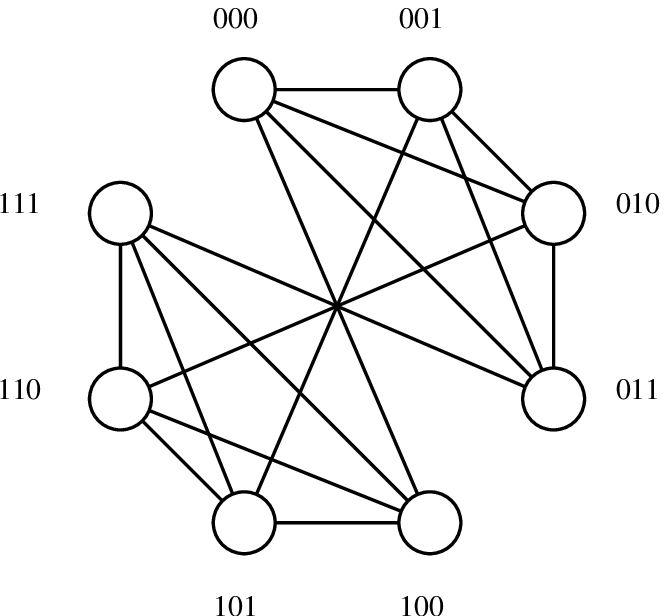}
}
\vspace{-1em}
\end{center}
\caption{(a) A 3-node directed graph. 
(b) The confusion graph corresponding to the integer tuple $(t_1, t_2, t_3) = (1,1,1)$. 
}
\label{fig:3-message}
\vspace{-1em}
\end{figure}

Given a directed graph $G$ and 
a length-$n$ integer tuple $\tv = (t_1, \ldots, t_n)$, the \emph{confusion graph}
$\G_{\tv}(G)$ is an undirected graph with $\prod_{i=1}^n 2^{t_i}$ vertices such that 
every vertex corresponds to a binary tuple $x^n$ and 
two vertices are connected iff the corresponding 
binary tuples are confusable. 
As an example, the confusion graph of the directed graph in Fig.~\ref{fig:3-message}(a) corresponding to $(t_1, t_2, t_3) = (1,1,1)$ is depicted in Fig.~\ref{fig:3-message}(b).

\begin{remark}
Every confusion graph is vertex transitive.
\end{remark}

The following lemma shows that given any directed graph $G$, 
confusion graph corresponding to a larger integer tuple has a larger fractional chromatic number.
The proof is presented in Appendix \ref{app:lemsleqt}.

\begin{lemma}
\label{lem:sleqt}
Let $\sv = (s_1, \ldots, s_n)$ and $\tv = (t_1, \ldots, t_n)$ be two integer tuples such that $\sv \leq \tv$. 
Then for any directed graph $G$ with $n$ vertices, we have
\begin{align}
\label{eq:sleqt}
\chi_f(\G_\sv(G)) \leq \chi_f(\G_\tv(G)).
\end{align}
\end{lemma}

\section{Problem Definitions}
\label{sec:def}

\subsection{Index Coding}
In the index coding problem, a sender wishes to communicate a tuple of $n$ messages, $x^n = (x_1, \ldots, x_n)$, $x_j \in \{0,1\}^{t_j}$, to their corresponding receivers using a shared noiseless channel.
Receiver $j \in [n]$ has prior knowledge of a subset  $x(A_j) := (x_i \suchthat i \in A_j)$,  $A_j \subseteq [n] \setminus \{j\}$, of the messages and wishes to recover $x_j$.
It is assumed that the sender is aware of $A_1, \ldots, A_n$.
The goal is to minimize the amount of information that should be broadcast from the sender to the receivers so that every receiver can recover its desired message.

Any instance of this problem, referred to collectively as the \emph{index coding problem}, 
is fully specified by the side information sets  $A_1, \ldots, A_n$. Equivalently, it can be specified
by a side information graph $G$ with $n$ nodes, in which
a directed edge $i \to j$ represents that receiver $j$ has message $i$ as side information ($i \in A_j$).
For instance, Fig.~\ref{fig:3-message}(a) shows the directed graph representing the index coding problem with $A_1 = \{2,3\}$, $A_2 = \{1\}$, and $A_3=\{1,2\}$.
We often identify an index coding problem with its side information graph and simply write ``index coding problem $G$.''

A $(t_1, \ldots, t_n, r)$ index code is defined by
\begin{itemize}
\item an encoder $\phi: \prod_{i=1}^n \{0,1\}^{t_i} \to \{0,1\}^r$ that maps $n$-tuple of messages $x^n$
to an $r$-bit index and
\item $n$ decoders $\psi_j: \{0,1\}^r \times \prod_{k \in A_j} \{0,1\}^{t_k} \to \{0,1\}^{t_j}$ that maps the received index $\phi(x^n)$ and the side information $x(A_j)$ back to $x_j$ for $j \in [n]$.
\end{itemize}
Thus, for every $x^n \in \prod_{i=1}^n \{0,1\}^{t_i}$,
\[
\psi_j(\phi(x^n), x(A_j)) = x_j, \quad j \in [n].
\]
A rate tuple $(R_1,\ldots,R_n)$ is said to be \emph{achievable} for the index coding problem $G$
if there exists a $(t_1, \ldots, t_n, r)$ index code such that 
\[
R_j \leq \frac{t_j}{r}, \quad j \in [n].
\]
The \emph{capacity region} $\Cr$ 
of the index coding problem is defined as the closure of the set of achievable rate tuples.
Let $\lambdav = (\lambda_1, \ldots, \lambda_n)$ be a non-negative real tuple.
Define the $\lambdav$-directed capacity $C(\lambdav)$ of the index coding problem $G$ as
\begin{align}
\label{eq:clambda}
C(\lambdav) = \max \{R \suchthat R \lambdav \in \Cr\}.
\end{align}

\begin{remark}
The capacity region can be written in terms of $\lambdav$-directed capacities.
\begin{equation}
\label{eq:lambdacapacity}
\Cr = \bigcup_{\lambdav \in \mathbb{R}_{\ge 0}^n \suchthat \sum_{i=1}^n \lambda_i = n} \left\lbrace\Rv \in \mathbb{R}_{\ge 0}^n \suchthat \Rv \leq C(\lambdav) \lambdav\right\rbrace.
\end{equation}
\end{remark}
If $\lambdav_1 = c \lambdav_2$ for some constant $c$, then $C(\lambdav_1) \lambdav_1 = C(\lambdav_2) \lambdav_2$ and thus in \eqref{eq:lambdacapacity}, it suffices to do the union over vectors $\lambdav$ such that $\sum_{i=1}^n \lambda_i = n$.
The $\ov$-directed capacity of the index coding problem $G$ is referred to as the symmetric capacity, 
\[ 
\csym = C(\ov) = \max \{R \suchthat (R, \ldots, R) \in \Cr\}.
\]
The reciprocal of the symmetric capacity is referred to as the broadcast rate $\beta$ of the index coding problem $G$.
For any nonnegative real vector $\muv = (\mu_1, \ldots, \mu_n)$, the $\muv$-weighted sum-capacity $\bar{C}(\muv)$ of the index coding problem $G$ is defined as
\[
\bar{C}(\muv) = \max\left\lbrace\sum_{i=1}^n \mu_i R_i \suchthat (R_1, \ldots, R_n) \in \Cr\right\rbrace.
\] 
The $\ov$-weighted sum-capacity is simply referred to as sum-capacity,
\[
\csum = \bar{C}(\ov) = \max\left\lbrace\sum_{i=1}^n R_i \suchthat (R_1, \ldots, R_n) \in \Cr\right\rbrace.
\]

\subsection{Locally Recoverable Distributed Storage}
In the locally recoverable distributed storage problem, data is to be stored in a network of $n$ servers such that by having access to all of the servers the stored data can be exactly recovered. 
Let $x_j \in \{0,1\}^{t_j}$ denote the content of server $j$, $j \in [n]$.
It is assumed that each server has access to the contents of a subset of the other servers $x(A_j)$, where $A_j \subseteq [n] \setminus \{j\}$ and  is referred to as the recoverability set of node $j$. 
The goal is to find the maximum amount of data that can be stored in the network such that 
if any of the servers fails, its content can be recovered from the contents of its recoverability set.
Any instance of the distributed storage problem is fully represented by the storage recovery graph $G=(V,E)$ in which each node represents a server and there exists a directed edge $i \to j$ iff server $i$ is in the recoverability set of server $j$. 

A $(t_1, \ldots, t_n, r)$ distributed storage code is defined by

\begin{itemize}
\item a message set $[2^r]$,
\item a one-to-one encoding function $x^n: [2^r] \to \prod_{i=1}^n \{0,1\}^{t_i}$ that assigns a codeword $x^n(m)$ to each message $m \in [2^r]$, (the set $\mathcal{C} = \{x^n(1), \ldots, x^n(2^r)\}$ is referred to as the codebook), and
\item $n$ recovery functions $f_j: \prod_{k \in A_j} \{0,1\}^{t_k} \to \{0,1\}^{t_j}$ that maps the contents of the recoverability set $x(A_j)$ to $x_j$ for $j \in [n]$. 
\end{itemize}
Thus, for every $x^n \in \mathcal{C}$,
\[
f(x(A_j)) = x_j, \quad j \in [n].
\]
A rate tuple $(R'_1,\ldots,R'_n)$ is said to be \emph{achievable} for the distributed storage problem $G$
if there exists a $(t_1, \ldots, t_n, r)$ distributed storage code such that 
\[
R'_j \geq \frac{t_j}{r}, \quad j \in [n].
\]
The \emph{optimal rate region} $\Rr$ 
of the distributed storage problem is defined as the closure of the set of achievable rate tuples.
For any non-negative real tuple $\lambdav = (\lambda_1, \ldots, \lambda_n)$, the $\lambdav$-directed optimal rate $R(\lambdav)$ of the distributed storage problem $G$  is defined as
\begin{align}
\label{eq:clambdap}
R(\lambdav) = \min \{R' \suchthat R' \lambdav \in \Rr\}.
\end{align}

\begin{remark}
The optimal rate region can be written in terms of $\lambdav$-directed optimal rates.
\begin{equation}
\label{eq:lambdacapacity-storage}
\Rr = \bigcup_{\lambdav \in \mathbb{R}_{\ge 0}^n \suchthat \sum_{i=1}^n \lambda_i = n} \left\lbrace \Rv' \in \mathbb{R}^n \suchthat \Rv' \geq R(\lambdav) \lambdav \right\rbrace.
\end{equation}
\end{remark}
The $\ov$-directed optimal rate of the distributed storage problem $G$ is referred to as the symmetric coding rate,
\begin{align*}
\rsym = R(\ov) = \min \{R' \suchthat (R', \ldots, R') \in \Rr \}.
\end{align*} 
The reciprocal of the symmetric coding rate is sometimes referred to as the normalized rate.
For any nonnegative real vector $\muv = (\mu_1, \ldots, \mu_n)$, the $\muv$-weighted optimal sum-rate $\bar{R}(\muv)$ of the distributed storage problem $G$ is defined as
\[
\bar{R}(\muv) = \min \left\lbrace\sum_{i=1}^n \mu_i R'_i \suchthat (R'_1, \ldots, R'_n) \in \Rr\right\rbrace.
\]
The $\ov$-weighted optimal sum-rate $\bar{R}(\ov)$ is simply referred to as optimal sum-rate
\begin{align*}
\rsum = \bar{R}(\ov) = \min \left\lbrace\sum_{i=1}^n R'_i \suchthat (R'_1, \ldots, R'_n) \in \Rr\right\rbrace.
\end{align*}

\subsection{Guessing Games on Directed Graphs}
Given a directed graph $G=(V,E)$, $V = [n]$, consider the following cooperative game among $n$ players.
Player $j \in [n]$ is associated to node $j$ and is assigned a value $x_j \in \{0,1\}^{t_j}$ independently from the other players.
It is assumed that player $j$ can observe the values assigned to a subset of the other players $x(A_j):= (x_i \suchthat i \in A_j)$, $A_j \subseteq [n] \setminus \{j\}$.
$A_j$ is referred to as the set of neighbors of player (node) $j$.
The players should simultaneously guess their own value and the goal is to maximize the probability that all players guess their value correctly.
No communication is allowed between the players, but they can agree on a strategy beforehand.
Note that the setting that we presented here is slightly different from the original one defined by Riis \cite{Riis2007} in that the range of the values assigned to the players can be different.

A $(t_1, \ldots, t_n, W)$ guessing strategy is defined by 
\begin{itemize}
\item $n$ functions $h_j: \prod_{k \in A_j} \{0,1\}^{t_k} \to \{0,1\}^{t_j}$ that maps the values of the neighbors $x(A_j)$ to $x_j$, for $j \in [n]$, and 
\item a set $W$ of $n$-tuples that can be guessed correctly using these functions,
\[
W := \bigcap_{j=1}^n \left\lbrace x^n \in \prod_{i=1}^n \{0,1\}^{t_i} \suchthat h_j(x(A_j)) = x_j\right\rbrace.
\]
\end{itemize}
Let $P_\textrm{win}$ be the probability of winning, namely, the probability that everyone guesses her value correctly.
If the players adopt a $(t_1, \ldots, t_n, W)$ strategy, i.e., player $j$, $j \in [n]$ uses function $h_j$ to guess her value based on the values of her neighbors, then  
\begin{align*}
P_\textrm{win} = |W|/\prod_{i=1}^n 2^{t_i}.
\end{align*}
Let $P_\textrm{rand}$ be the probability of winning if 
every player guesses her value randomly.
As  player $j$, $j \in [n]$, is correct with probability $\frac{1}{2^{t_j}}$ independent of others, we have
\[
P_\textrm{rand} = 1/\prod_{i=1}^n 2^{t_i}.
\]
To measure the performance of adopting a strategy, in \cite{Riis2007} guessing number is defined as the logarithm of the ratio between the winning probabilities of an strategy and a random guess.
In the following, we define the guessing number for the general case where the range of the values assigned to the players are different.

\begin{definition}
Given a directed graph $G$, the guessing number of a $(t_1, \ldots, t_n, W)$ guessing strategy is defined as
\begin{align}
\label{eq:guessing-number}
k(G,\tv) = \log_s \left( P_\textrm{win}/P_\textrm{rand} \right),
\end{align}
where $s = 2^{\frac{1}{n}\sum_{i=1}^n t_i}$,
and $P_\textrm{win}$ is the probability that the players win if they adopt that strategy.
\end{definition}%
Note that for the case where $t_j = t, \forall j$, we have 
$k(G, t) = \log(|W|)/t$.
The optimal guessing number $k(G)$ of a directed graph $G$ is defined as
\begin{align}
\label{eq:guessing-number-optimal}
k(G) = \sup_\tv \sup k(G,\tv),
\end{align}
where the second supremum is over all $(t_1, \ldots, t_n, W)$ strategies.
The following is an alternative way to measure the performance of adopting a strategy.
\begin{definition}
Given a directed graph $G$, the complementary guessing number of a $(t_1, \ldots, t_n, W)$ guessing strategy is defined as
\begin{align}
\label{eq:guessing-number-comp}
k'(G,\tv) = \log_s \left( 1/P_\textrm{win} \right),
\end{align}
where $s = 2^{\frac{1}{n}\sum_{i=1}^n t_i}$ and $P_\textrm{win}$ is the probability that the players win if they adopt that strategy.
\end{definition}
The optimal complementary guessing number is defined in a similar way.
\begin{align}
\label{eq:guessing-number-optimal-comp}
k'(G) = \inf_\tv \inf k'(G,\tv),
\end{align}
where the second infimum is over all $(t_1, \ldots, t_n, W)$ guessing strategies.

\begin{remark}
For any $(t_1, \ldots, t_n, W)$  guessing strategy, we have
$k'(G,\tv) = n - k(G,\tv)$,
and thus
\begin{align}
\label{eq:k-k'}
k'(G) = n - k(G).
\end{align}
\end{remark}

Based on the definitions, it is easy to see the following relationship between the guessing games and the distributed storage problem.
\begin{remark}
Given any directed graph $G$, a $(t_1, \ldots, t_n, r)$ distributed storage code exists iff a $(t_1, \ldots, t_n, W)$ guessing strategy with $|W| \geq 2^r$ exists. 
\end{remark}
We will return to the relationship between these two problems at the and of Section \ref{sec:duality}.

\section{Capacity Region, Optimal Rate Region, and Guessing Number via Confusion Graph}
\label{sec:capacity}

In order to shed light on how these three problems are related,
we will review the characterization of the capacity region of the index coding problem, the optimal rate region of the distributed storage problem, and the guessing number of  guessing games on graphs via the notion of confusion graph.

\subsection{Index Coding}
In \cite{Arbabjolfaei--Kim2015}, the capacity region is characterized through the fractional chromatic number of the confusion graph.

\begin{proposition}
\label{prop:graphregion3}
The capacity region $\Cr$ of the index coding problem $G$ is the closure
of all rate tuples $(R_1, \ldots, R_n)$ such that
\begin{equation}
\label{eq:graphregion3}
R_j \leq \frac{t_j}{\log(\chi_f(\G_{\tv}(G)))}, \quad j \in [n],
\end{equation}
for some $\tv = (t_1, \ldots, t_n)$.
\end{proposition}

\subsection{Distributed Storage and Guessing Game}

In \cite{Mazumdar2014}, Mazumdar characterized the normalized rate of the distributed storage problem through the independence number of the confusion graph. The following is a simple generalization of Mazumdar's result.

\begin{proposition}
\label{prop:diststorageconfusion}
A rate tuple $(R'_1, \ldots, R'_n)$ is achievable for the distributed storage problem $G$ iff there exists an integer tuple $\tv = (t_1, \ldots, t_n)$ such that
\begin{align*}
R'_j \geq \frac{t_j}{\floor{\log(\alpha(\G_\tv(G)))}}, \quad j \in [n].
\end{align*}
\end{proposition}

\begin{IEEEproof} \emph{Sufficiency (achievability).} 
For a given tuple $\tv = (t_1, \ldots, t_n)$, consider a maximal independent set of the confusion graph 
$\G = \G_{\tv}(G)$.
By the definition of the confusion graph, no two $n$-tuples in this independent set are confusable
and therefore for these $\alpha(\G)$ $n$-tuples, contents of each server is a function of the contents of its recoverability set.
Therefore, it is possible to use these $\alpha(\G)$ $n$-tuples to store $r = \floor{\log(\alpha(\G))}$ bits in the distributed network. 
This proves the existence of a $(t_1, \ldots, t_n, \floor{\log(\alpha(\G_\tv(G)))})$ distributed storage code.

\emph{Necessity (converse).} 
Consider any $(t_1, \ldots, t_n, r)$ distributed storage code, which has at least $2^r$ distinct $n$-tuples that satisfy the required function relationship.
By definition, these $n$-tuples form an independent set of the
confusion graph $\G = \G_\tv(G)$. 
Thus, $\alpha(\G) \ge 2^r$, or equivalently, 
$r \le \floor{\log(\alpha(\G))}$.
Therefore, any achievable $(R'_1, \ldots, R'_n)$ must satisfy
\[
R'_j \ge \frac{t_j}{\floor{\log(\alpha(\G_\tv(G)))}}, \quad j \in [n],
\]
for some $\tv = (t_1,\ldots,t_n)$.
\end{IEEEproof}

Since
\begin{align*}
\frac{kt_j}{\floor{k\log(\alpha(\G_\tv(G)))}} \le \frac{kt_j}{k\log(\alpha(\G_\tv(G)))-1},
\end{align*}
letting $k \to \infty$ establishes the following.

\begin{proposition}
\label{prop:cap-dist-storage}
The optimal rate region $\Rr$ of the distributed storage problem $G$ is the closure of all rate tuples $(R'_1, \ldots, R'_n)$ such that
\begin{align*}
R'_j \ge \frac{t_j}{\log(\alpha(\G_\tv))}, \quad j \in [n],
\end{align*}
for some $\tv = (t_1, \ldots, t_n)$.
\end{proposition}

\medskip

Using an argument similar to the proof of Proposition \ref{prop:diststorageconfusion}, the optimal guessing number of the guessing game on directed graph $G$ can be characterized via the independence number of confusion graphs.
\begin{proposition}
\label{prop:k-guessing-game}
For the guessing game on directed graph $G$ on $n$ nodes we have
\begin{align*}
k(G) = \sup_{\tv \in \mathbb{Z}_{\geq 0}^n} ~\log(\alpha(\G_\tv))/(\frac{1}{n}\sum_{i=1}^n t_i).
\end{align*}
\end{proposition}

\section{Relationships}
\label{sec:duality}
In this section, we first discuss the complementarity between the index coding and the distributed storage and then clarify the relationship between guessing games and the distributed storage problem.

\subsection{Index Coding Versus Distributed Storage}

For any length-$n$ integer tuple $\tv$, the confusion graph $\G_\tv$ is vertex transitive. Therefore, Lemma \ref{lem:vertextransitive} yields
\begin{align}
\label{eq:chivsalpha}
\log(\chi_f(\G_\tv)) = \sum_{i=1}^n t_i - \log(\alpha(\G_\tv)),
\end{align}
which is the key to establishing the complementarity between the index coding problem and the distributed storage problem.
We start by stating the following proposition.
The proof of the proposition is relegated to  Appendix \ref{app:propClambdarational}.

\begin{proposition}
\label{prop:Clambdarational}
For any directed graph $G$ on $n$ nodes and any $\lambdav \in \mathbb{Q}_{\geq 0}^n$, 
\begin{align}
\label{eq:Clambda}
C(\lambdav) &= \sup_{r \suchthat r\lambdav \in \mathbb{Z}_{\geq 0}^n} \frac{r}{\log(\chi_f(\G_{r\lambdav}(G)))},\\
\label{eq:Cplambda}
R(\lambdav) &= \inf_{r \suchthat r\lambdav \in \mathbb{Z}_{\geq 0}^n} \frac{r}{\log(\alpha(\G_{r\lambdav}(G)))}.
\end{align}
\end{proposition}

The following theorem is the main result of this paper which establishes the complementarity between the $\lambdav$-directed capacity $C(\lambdav)$ and the $\lambdav$-directed optimal rate $R(\lambdav)$, for any nonnegative real tuple $\lambdav$.
\begin{theorem}
\label{thm:duality}
For any directed graph $G$ on $n$ nodes and any $\lambdav \in \mathbb{R}_{\geq 0}^n$, we have
\vspace{-1em}
\begin{align}
\label{eq:duality}
\frac{1}{C(\lambdav)} = \sum_{i=1}^n \lambda_i - \frac{1}{R(\lambdav)}.
\end{align}
\end{theorem}

\begin{IEEEproof}
For $\lambdav \in \mathbb{Q}_{\geq 0}^n$, we have
\begin{align}
\label{eq:duality1}
C(\lambdav) &= \sup_{r \suchthat r\lambdav \in \mathbb{Z}_{\geq 0}^n} \frac{r}{\log(\chi_f(\G_{r\lambdav}(G)))} \\
\label{eq:duality2}
&= \sup_{r \suchthat r\lambdav \in \mathbb{Z}_{\geq 0}^n}  \frac{r}{r\sum_{i=1}^n \lambda_i - \log(\alpha(\G_{r\lambdav}(G)))} \\
\label{eq:duality3}
&= \frac{1}{\sum_{i=1}^n \lambda_i - \frac{1}{R(\lambdav)}},
\end{align}
where \eqref{eq:duality1} and \eqref{eq:duality3} follow from Proposition \ref{prop:Clambdarational}, and \eqref{eq:duality2} follows from \eqref{eq:chivsalpha}.

For $\lambdav \not \in \mathbb{Q}^n$, \eqref{eq:duality} holds due to the continuity of the functions $C(\lambdav)$ and $R(\lambdav)$ and $\mathbb{Q}$ being dense in $\mathbb{R}$.
\end{IEEEproof}

Due to equations \eqref{eq:lambdacapacity} and \eqref{eq:lambdacapacity-storage}, 
the above theorem establishes the complementarity between the two problems in the strong sense that given the capacity region of the index coding problem $G$ (more precisely, given the boundary points of the capacity region), Theorem \ref{thm:duality} completely determines the optimal rate region for the distributed storage problem $G$ and vice versa.
This includes as an special case the complementarity between the symmetric capacity of the index coding problem and the symmetric coding rate of the distributed storage established by Mazumdar \cite{Mazumdar2014}, and by Shanmugam and Dimakis \cite{Shanmugam--Dimakis2014}.
\begin{corollary}
Setting $\lambdav = \ov$ in Theorem \ref{thm:duality} yields
\begin{align}
\frac{1}{\csym} = n - \frac{1}{\rsym}.
\end{align}
\end{corollary}

Equation \eqref{eq:chivsalpha} can also be used to show how the sum-capacity of the index coding problem is related to the optimal sum-rate of the distributed storage problem.
The proof is straightforward and is omitted.
\begin{theorem}
\label{thm:sumdual}
\begin{align}
\frac{1}{\csum} = 1 - \frac{1}{\rsum}.
\end{align}
\end{theorem}


\subsection{Guessing Games Versus Distributed Storage}
The following proposition shows that for any directed graph $G$, the optimal guessing number of the guessing game on $G$ is inversely related to the optimal sum-rate of the distributed storage problem $G$.
\begin{theorem}
\label{thm:ggvsds}
For any directed graph $G$ on $n$ nodes
\begin{align*}
k(G) = \frac{n}{\rsum}.
\end{align*}
\end{theorem}

\begin{IEEEproof} For any tuple $\tv$, let $s(\tv) = 2^{\frac{1}{n}\sum_{i=1}^n t_i}$, then
\begin{align}
\rsum &= \min_{R' \in \Rr} \sum_{i=1}^n R'_i \nonumber \\
\label{eq:ggvsds1}
&= \inf_{\tv} \frac{\sum_{i=1}^n t_i}{\log(\alpha(\G_\tv))}  \\
&= \frac{n}{\sup_{\tv} \frac{\log(\alpha(\G_\tv))}{\log(s(\tv))}} 
= \frac{n}{k(G)},
\label{eq:ggvsds2}
\end{align}
where \eqref{eq:ggvsds1} follows from Proposition \ref{prop:cap-dist-storage}, and \eqref{eq:ggvsds2} follows from Proposition \ref{prop:k-guessing-game}.
\end{IEEEproof}

Combining Theorems \ref{thm:sumdual} and \ref{thm:ggvsds}, and \eqref{eq:k-k'} yields the inverse relationship between the optimal complementary guessing number and the index coding sum-capacity.
\begin{corollary}
For any directed graph $G$ on $n$ nodes
\begin{align*}
k'(G) = \frac{n}{\csum}.
\end{align*}
\end{corollary}

\section{Concluding remarks}
\label{sec:conclusion}
In this paper, we surveyed three problems and elaborated on the connections between their optimal solutions.
This can be used in directly translating any achievability scheme or converse result for one problem to the other problems. 
For example, in \cite{Baber--Christofides--Dang--Riis--Vaughan2013}, it is shown that even for undirected graphs, the guessing strategy based on the fractional clique covering \cite{Christofides--Markström2011} is not optimal.
It is also shown via an example that non-Shannon inequalities provide a better bound on the guessing number of undirected graphs than Shannon inequalities.
The  established connections between the problems answer the corresponding open problems in the context of index coding, namely, even when restricted to undirected graphs, neither the fractional clique covering inner bound nor the polymatroidal outer bound is tight for the index coding problem.

\section{Acknowledgments}
This work was supported by the National Science Foundation under Grant CCF-1320895
and the Korean Ministry of Science, ICT and Future Planning under the
Institute for Information and Communications Technology Promotion Grant B0132-15-1005
(Development of Wired-Wireless Converged 5G Core Technologies).

\appendices

\section{Proof of Lemma \ref{lem:sleqt}}
\label{app:lemsleqt}

First assume that $s_j + k = t_j$ for some $j \in [n]$ and some positive integer $k$ and $s_i = t_i, \forall i \not = j$.  
In this case, we will prove the lemma  by contradiction.
Assume that \eqref{eq:sleqt} does not hold. 
Then as any confusion graph is vertex transitive, by Lemma \ref{lem:vertextransitive}, we have
\begin{align}
\label{eq:cont-assumption}
\alpha(\G_\tv(G)) > 2^k \alpha(\G_\sv(G)).
\end{align}
Each vertex of $\G_\tv$ is associated with an $n$-tuple that has $t_i$ bits for user $i \in [n]$.
Consider the $\alpha(\G_\tv)$ $n$-tuples in a maximal independent set of $\G_\tv$ and partition them into (at most $2^k$) subsets based on the first $k$ bits of user $j$.
As these $k$ bits are the same for all the members of each partition, after removing these $k$ bits from all the $n$-tuples, each partition will correspond to an independent set of $\G_\sv$.
However, there are at most $2^k$ partitions and hence if \eqref{eq:cont-assumption} holds, due to the pigeonhole principle, there exists a partition with more than $\alpha(\G_\sv)$ members, i.e., there exists an independent set of size more than $\alpha(\G_\sv)$ in $\G_\sv$, which contradicts the definition of the independence number of a graph. Therefore, \eqref{eq:sleqt} holds if the two integer tuples differ only at one element. Applying this (at most $n$ times) to length-$n$ tuples that differ only at one element, completes the proof of the lemma. 

\section{Proof of Proposition \ref{prop:Clambdarational}}
\label{app:propClambdarational}

Due to Proposition \ref{prop:graphregion3}, 
$
\Cr = cl(\Cr^{\degree})
$,
where
\begin{equation*}
\Cr^{\degree} = \{\Rv \in \mathbb{R}^n_{\geq 0} \suchthat \Rv \leq \frac{\tv}{\log(\chi_f(\G_\tv))} ~\text{for some}~ \tv \in \mathbb{Z}^n_{\geq 0} \}.
\end{equation*}
The following lemma shows that the $\lambdav$-directed capacity defined in \eqref{eq:clambda}, can also be defined in terms of $\Cr^{\degree}$.
\begin{lemma}
\label{lem:supeqmax}
For any non-negative real tuple $\lambdav$,
\begin{align}
C(\lambdav) = \sup \{R \suchthat R \lambdav \in \Crn\}.
\end{align}
\end{lemma}

\begin{IEEEproof}
Let $R^* = \sup \{R \suchthat R \lambdav \in \Crn\}$, then $R^* \lambdav \in \Cr$ and by the definition of $C(\lambdav)$ we have
$R^* \leq C(\lambdav)$. 

Assume that $R^* < C(\lambdav)$.
Let \[ \epsilon = \frac{1}{2} (C(\lambdav)-R^*) \min_{i \suchthat \lambda_i > 0} \lambda_i,\]
and define the $\epsilon$-neighborhood  $N_\epsilon(C(\lambdav) \lambdav)$ as
\[
N_\epsilon(C(\lambdav) \lambdav) = \bigcap_{i \suchthat \lambda_i > 0} \left\lbrace\Rv \in \mathbb{R}^n \suchthat e_i^T (C(\lambdav) \lambdav - \Rv) < \epsilon\right\rbrace,
\]
where all  of the  components of the $n \times 1$  vector $e_i$ are  zero,  except  the $i$-th component, which is one.
If $N_\epsilon(C(\lambdav) \lambdav) \cap \Crn = \emptyset$, then it contradicts the fact that $C(\lambdav) \lambdav$ belongs to $\Cr$.
Alternatively, if $N_\epsilon(C(\lambdav) \lambdav) \cap \Crn \not = \emptyset$, then there exists $R > R^*$ such that $R\lambdav \in \Crn$, which contradicts the definition of $R^*$. 
Therefore, $R^* = C(\lambdav)$ and the proof is complete. 
\end{IEEEproof}

\begin{IEEEproof}[Proof of Proposition \ref{prop:Clambdarational}]
Let $\lambda = (\frac{a_1}{b}, \ldots, \frac{a_n}{b})^T$, $b \in \mathbb{N}$, and  $a_1, \ldots, a_n \in \mathbb{Z}_{\geq 0}$. 
If $r\lambdav \in \mathbb{Z}^n_{\geq 0}$, then due to Proposition \ref{prop:graphregion3}, we have
\vspace{-0.3em}
\[
\frac{r \lambdav}{\log(\chi_f(\G_{r\lambdav}(G)))} \in \Cr.
\]
Therefore, we have
\[
C(\lambdav) \geq \sup_{r \suchthat r\lambdav \in \mathbb{Z}^n_{\geq 0}} \frac{r}{\log(\chi_f(\G_{r\lambdav}(G)))}.
\]
Next, let $R$ be any real number such that $R\lambda \in \Cr^0$. 
Then, there exists integer tuple $\tv$ such that
$R\lambdav \leq \tv/\log(\chi_f(\G_\tv))$, 
and hence $R \leq \frac{t_i}{\lambda_i}/\log(\chi_f(\G_\tv))$, for all $i$ such that $\lambda_i > 0$.
Let 
\vspace{-0.5em}
\begin{align}
\label{eq:argmin}
j = \arg\min_{i \suchthat \lambda_i > 0} \frac{t_i}{\lambda_i},
\end{align}
then we have 
\vspace{-0.5em}
\begin{align}
\label{eq:appendix1}
R \leq \frac{q}{a_j \log(\chi_f(\G_\tv))},
\end{align}
where $q = t_jb$. 
Due to \eqref{eq:argmin}, $a_j \tv \geq q\lambdav = t_j(a_1, \ldots, a_n)^T \in \mathbb{Z}_{\geq 0}^n$ and we have
\begin{align}
\label{eq:ineq1}
\log(\chi_f(\G_{q \lambda})) &\leq \log(\chi_f(\G_{a_j \tv})) \\
\label{eq:ineq2}
&\leq \log(\chi_f(\G_\tv^{a_j})) \\
\label{eq:ineq3}
&= a_j \log(\chi_f(\G_\tv)),
\end{align}
where \eqref{eq:ineq1} follows from Lemma \ref{lem:sleqt}, \eqref{eq:ineq2} follows from the fact that the set of edges of $\G_{a_j \tv}$ is a subset of the set of edges of $\G_\tv^{a_j}$, and \eqref{eq:ineq3} follows from Lemma \ref{lem:orcoloring}. 
Combining \eqref{eq:appendix1} and \eqref{eq:ineq3}, we have
\[
R \leq \frac{q}{\log(\chi_f(\G_{q \lambda}))} \leq \sup_{r \suchthat r\lambdav \in \mathbb{Z}_{\geq 0}^n} \frac{r}{\log(\chi_f(\G_{r\lambdav}(G)))},
\]
which together with Lemma \ref{lem:supeqmax} yields
\begin{align*}
C(\lambdav) \leq  \sup_{r \suchthat r\lambdav \in \mathbb{Z}_{\geq 0}^n} \frac{r}{\log(\chi_f(\G_{r\lambdav}(G)))},
\end{align*} 
and hence \eqref{eq:Clambda} holds.
Following similar steps as above, one can show that \eqref{eq:Cplambda} also holds.
\end{IEEEproof}

\bibliographystyle{IEEEtran}
\bibliography{nit} 

\end{document}